\documentclass{aa}

\usepackage{amssymb}
\usepackage{amsmath}
\usepackage{graphicx}
\usepackage{natbib}
\usepackage[varg]{txfonts}
\usepackage{hyperref}

\hypersetup{
    colorlinks=true,
    linkcolor=blue,
    filecolor=magenta,
    urlcolor=blue,
    citecolor=blue,
}

\bibpunct{(}{)}{;}{a}{}{,}
\graphicspath{{figs/},}

\newcommand{\kev}{\ensuremath{\,\text{keV}}\xspace}
\newcommand{\cyc}{\ensuremath{\text{cyc}}}

\newcommand{\diff}{\ensuremath{\mathrm{d}}}
\newcommand{\gs}{\,\mathrm{g}\,\mathrm{s}^{-1}}

\newcommand{\gauss}{\,\mathrm{G}}

\newcommand{\el}{\mathrm{e}}

\begin{document}

\title{Vacuum polarization alters the spectra of accreting X-ray pulsars}

\author{
        E.~Sokolova-Lapa\inst{\ref{inst1}
        \thanks{\email{ekaterina.sokolova-lapa@fau.de}}}
        \and J.~Stierhof\inst{\ref{inst1}}
        \and T.~Dauser\inst{\ref{inst1}}
        \and J.~Wilms\inst{\ref{inst1}}
}

\institute{Dr.~Karl Remeis-Observatory and Erlangen Centre for
  Astroparticle Physics, Friedrich-Alexander
  Universit\"at Erlangen-N\"urnberg,
Sternwartstr.~7, 96049 Bamberg, Germany\label{inst1}
}

\abstract{It is a common belief that for magnetic fields typical for
  accreting neutron stars in High-Mass X-ray Binaries vacuum
  polarization only affects the propagation of polarized emission in
  the neutron star magnetosphere. We show that vacuum resonances can
  significantly alter the emission from the poles of accreting neutron
  stars. The effect is similar to vacuum polarization in the
  atmospheres of isolated neutron stars and can result in suppression
  of the continuum and the cyclotron lines. It is enhanced by magnetic
  Comptonization in the hot plasma and proximity to the electron
  cyclotron resonance. We present several models to illustrate the
  vacuum polarization effect for various optically thick media and
  discuss how the choice of polarization modes affects the properties
  of the emergent radiation by simulating polarized energy- and
  angle-dependent radiative transfer. Polarization effects,
  including vacuum polarization, crucially alter the emission
  properties. Together with strongly angle- and energy-dependent
  magnetic Comptonization, they result in a complex spectral shape,
  which can be described by dips and humps on top of a power-law-like
  continuum with high-energy cutoff. These effects provide a possible
  explanation for the common necessity of additional broad Gaussian
  components and two-component Comptonization models that are used to
  describe spectra of accreting X-ray pulsars. We also demonstrate the
  character of depolarization introduced by the radiation field's
  propagation inside the inhomogeneous emission region. }


\date{Received ... / Accepted ... }

\maketitle

\section{Introduction}\label{sec:intro}

The formation of the X-ray spectra of accreting highly magnetized
($B\sim 10^{12}$\,G) neutron stars in High-Mass X-ray Binaries (HMXBs)
is a complex process. The emitted continuum is mainly shaped by
magnetic Comptonization in an accretion column or accretion mound.
Resonant Compton scattering also results in the appearance of
absorption line-like features in the spectra, or ``cyclotron lines''
\citep{staubert2019}. Recent observations have revealed the complexity
of continuum shapes, which often cannot be described only by a simple
power law with high-energy cutoff \citep{mueller2013},
but require additional broad Gaussian features in absorption or
emission at intermediate energies \citep[also known as the
``10-keV'' feature; see, e.g.,][]{ferrigno2009, diez2022}.
Alternatively, the continuum can be described as the sum of two
Comptonization spectra from plasmas with different temperatures
\citep[see, e.g.,][]{doroshenko2020a}. In contrast, the available
angle- and polarization-averaged models for continuum formation
\citep[e.g.,][]{becker2007,west2017a,west2017b} yield a smooth spectral shape.
The rich observational background therefore remains ahead of the
spectral modeling, calling for new emission models that incorporate
further physics, such as the effects of polarization and angular
redistribution.

Polarization of radiation in the strong magnetic field introduces
further complexity in the modeling. Apart from the magnetoactive
plasma, photon polarization states can be affected by the quantum
electrodynamic (QED) effect known as vacuum polarization. For the
cold, absorption-dominated inhomogeneous atmospheres of highly magnetized isolated
neutron stars ($B\sim10^{14}\gauss$), radiative transfer solutions
were obtained \citep[e.g.,][]{lai2002,ho2003,oezel2003},
some of which also included partial mode conversion
\citep{ho2004,adelsberg2006}.
For the lower fields characteristic of accreting neutron stars in
HMXBs, however, the vacuum resonance has not been included in the
radiative transfer discussions of the last three decades. Early
studies of vacuum polarization effects by \citet{pavlov1979},
\citet{ventura1979b}, \citet{kaminker1982}, and \citet{soffel1985}
addressed the importance of this phenomenon for homogeneous media
only, without Compton scattering,
finding that in this case the vacuum resonance results in a narrow
low-energy absorption feature.
\citet{meszaros1985a,meszaros1985b} noted the importance of the
combined vacuum and plasma polarization modes for the shape of the
cyclotron line profile based on their homogeneous models including
Comptonization, but no qualitative comparison with other cases
was made.
Later models for the accretion column and atmospheric emission
considered only the polarization-averaged case \citep[see,
e.g.,][]{becker2007, farinelli2016, west2017b} or used pure magnetized
plasma \citep{mushtukov2021, caiazzo2021b, caiazzo2021a} and pure
vacuum normal modes \citep{sokolova-lapa2021}. Models for cyclotron
lines on top of a predefined continuum also largely focused on the
pure vacuum case \citep[see, e.g.,][]{araya1999,schwarm2017a,schwarm2017b}.

Previous claims that for $B\sim10^{12}\gauss$ vacuum polarization does
not alter the emission spectra \citep{ho2004, adelsberg2006} should be
taken with a grain of salt, as they were based on computations for
cool absorption-dominated atmospheres. Vacuum polarization in the hot
Comptonizing medium of accreting highly-magnetized neutron stars thus
remains poorly studied. Here, we would like to highlight the effect of
vacuum polarization in these environments and demonstrate its possible
influence on the spectral shape and polarization signal. We perform a
comparison for different choices of photon normal modes. We introduce
the problem of the mode definition in Sect.~\ref{sec:modes}. In
Sect.~\ref{sec:model}, we present several models for emission from
homogeneous and inhomogeneous media.
Section~\ref{sec:concl} provides the discussion and conclusion of this
study.

\section{Polarization modes}
\label{sec:modes}

The highly magnetized plasma in the vicinity of a neutron star is a
complex medium, which can support various types of waves. In the
medium, transverse electromagnetic waves, with electric vector,
$\vec{E}$, perpendicular to the wave vector, $\vec{k}$, are allowed to
propagate at frequencies higher than the electron plasma frequency.
For modeling the X-ray emission from accreting neutron stars in HMXBs,
the description of the radiation field in terms of two transverse
waves, the two polarization normal modes, became a standard approach.
We outline the classical definition of the two modes in
Sect.~\ref{sec:modclas} and discuss their ambiguity when both, the
plasma and the QED vacuum alter the radiation propagation in
Sect.~\ref{sec:ambig}.

\subsection{Mode definition and ellipticity}
\label{sec:modclas}

\begin{figure}
  \centering
    \resizebox{!}{10\baselineskip}{\includegraphics{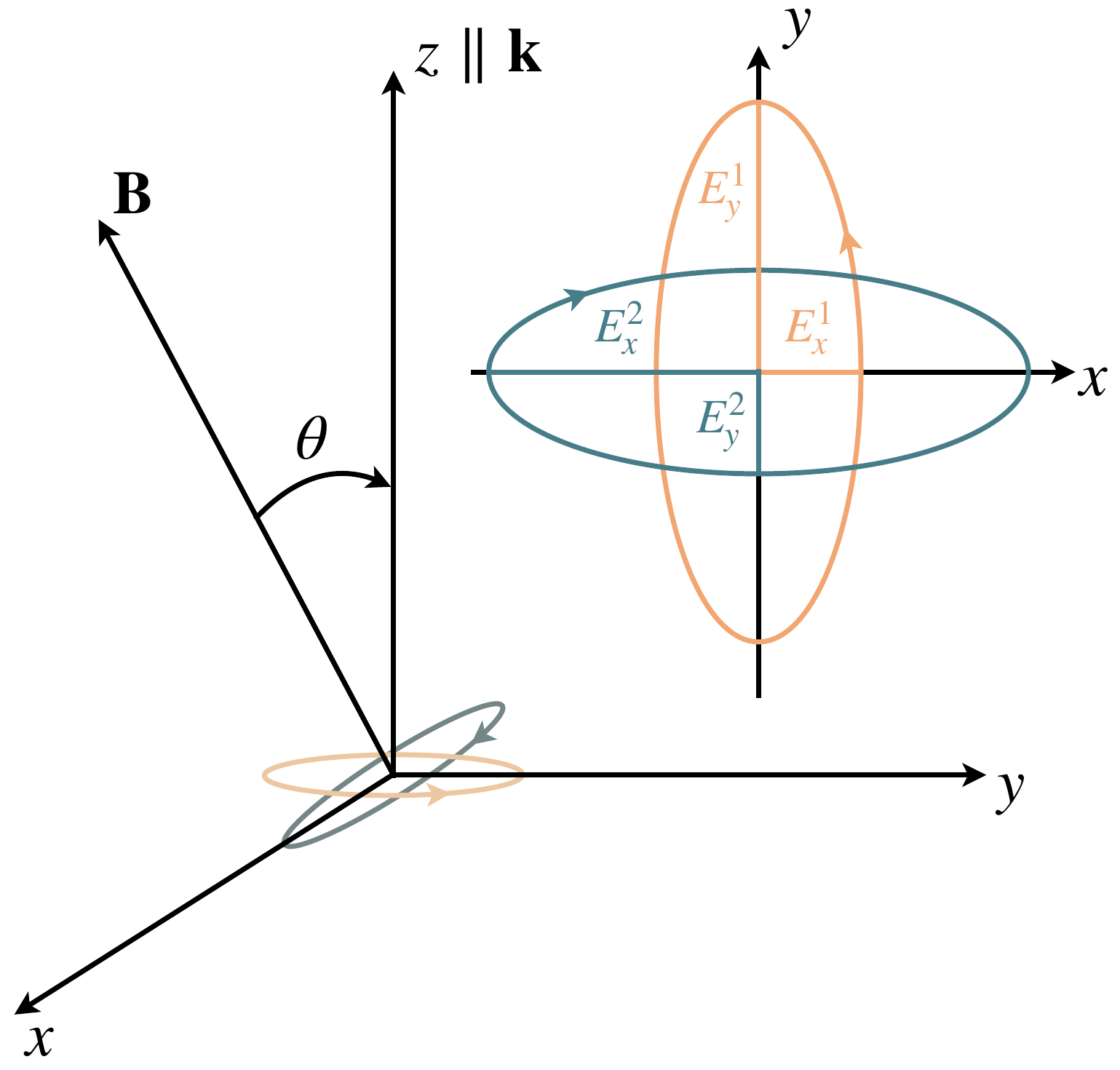}}
    \resizebox{!}{10\baselineskip}{\includegraphics{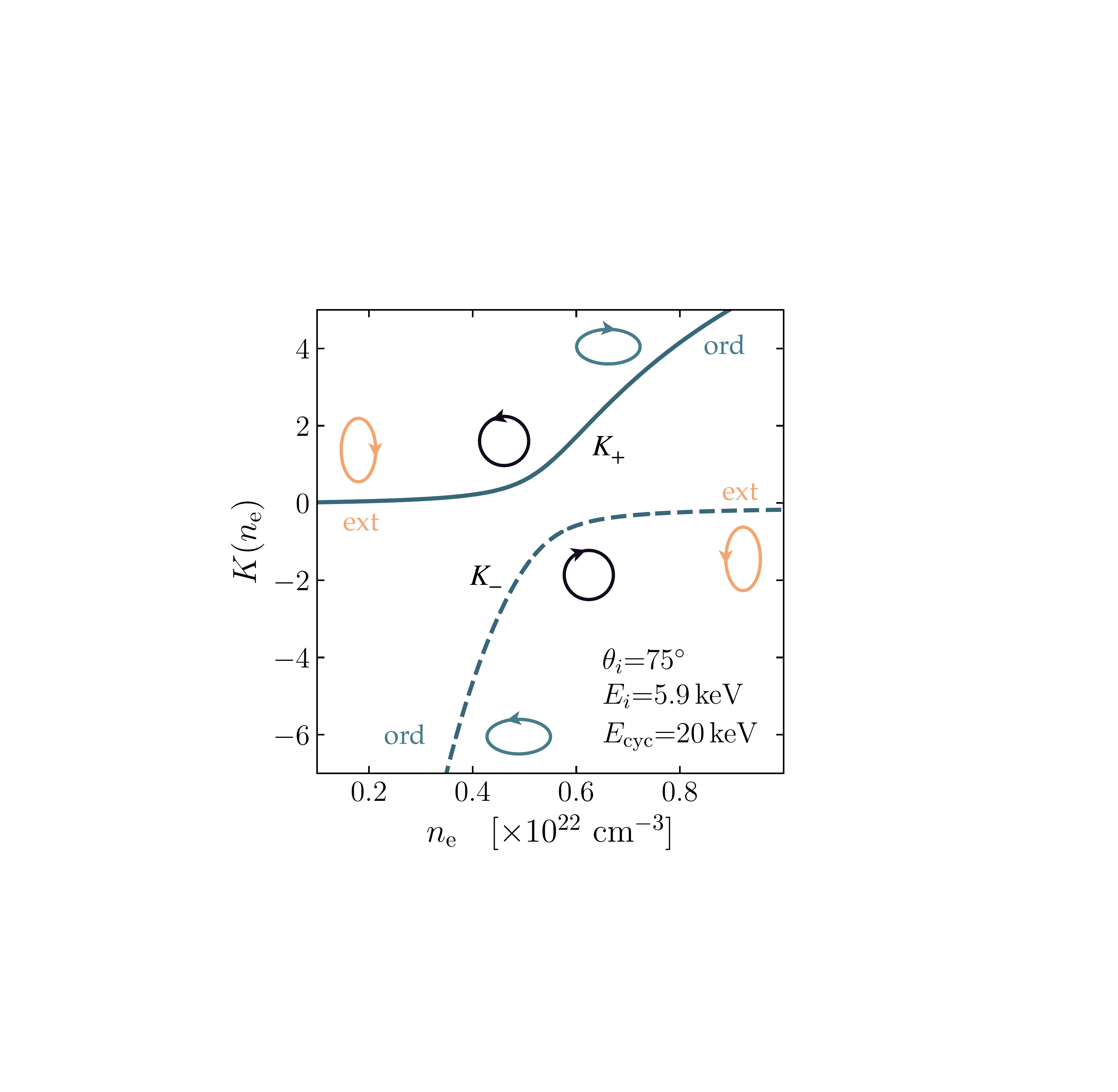}}
    \caption{\emph{Left:} Coordinate system chosen to define ellipses
      of the polarization modes. \emph{Right:} Dependency of the
      ellipticity of the polarization modes on the electron number
      density. The vacuum resonance occurs in the center of the
      figure. Based on Fig.~1 from \citet{lai2003}.}
    \label{fig:modes}
\end{figure}

For high-energy radiation in a magnetoactive plasma with
$B\sim 10^{12}$\,G, the electron response governs wave propagation.
The solution of the wave equation with a linear response in the form
of two independent modes is known for different temperature regimes
\citep[e.g.,][]{ventura1979a,pavlov1980,nagel1981b}. The polarization
vectors, $\vec{E}$, of the two normal modes have opposite directions
of rotation. The modes are elliptically polarized, where the
ellipticity $K=E_{x}/E_{y}$ in a coordinate system in which the
$z$-axis is parallel to $\vec{k}$ (Fig.~\ref{fig:modes}, left). The
``ordinary'' mode is defined as the left-hand polarized wave, with
$\vec{E}$ oscillating mainly in the ($\vec{B}$, $\vec{k}$)-plane
($|K|\gg1$), where $\vec{B}$ is the external magnetic field vector. The
polarization vector of the right-hand polarized ``extraordinary'' mode
oscillates mainly perpendicularly to this plane ($|K|\ll1$). For the
extraordinary mode, the co-directional rotation of $\vec{E}$ with
electrons in the external $B$-field allows the mode to resonate
at the cyclotron frequency. In the absence of damping, the ellipses of
the two modes are of equal shape, perpendicular to each other.

The effect of vacuum polarization in a strong $B$-field alters
this picture due to the birefringence introduced by QED's virtual
electron-positron plasma, which allows both modes to exhibit cyclotron
resonances.
In the absence of a real plasma, the equality of the virtual charges
makes the normal modes linear and strictly orthogonal to each other
(except for $\vec{k} \parallel\vec{B}$).
The combined effect of the virtual and the real
plasma results in a more complex behavior of the modes,
causing occasional departure from orthogonality.

\subsection{Mode ambiguity}
\label{sec:ambig}

\begin{figure}
\centering
     \resizebox{\hsize}{!}{\includegraphics{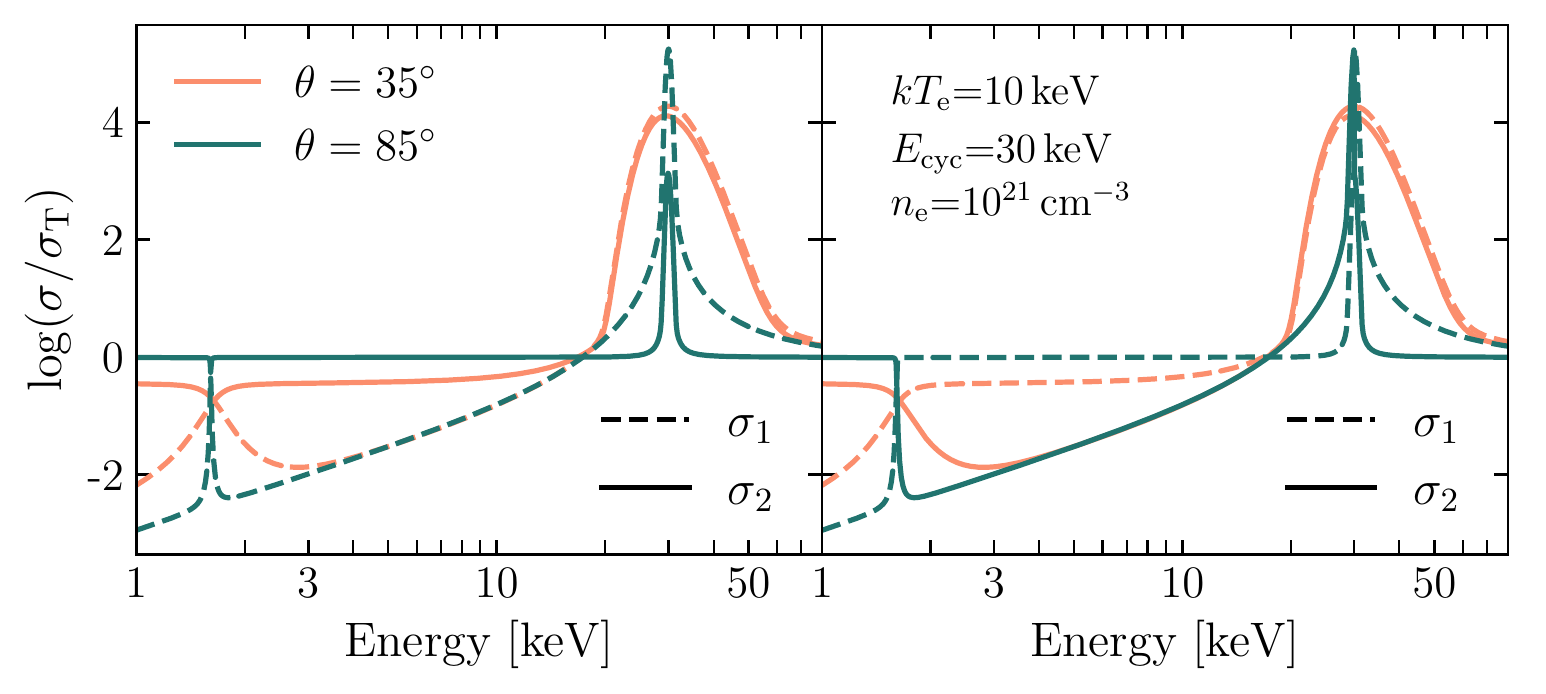}}
     \caption{Cross sections for magnetic Compton scattering for two
       different propagation angles, $\theta$. The modes are shown
       \emph{left}: for the case of discontinuous and \emph{right}:
       continuous refractive indices for $B\approx2.6\times10^{12}\gauss$.
     }
    \label{fig:crs}
\end{figure}

For a given $\vec{B}$-field, a given electron number density of the
plasma, $n_\el$, and a photon propagation angle, $\theta$, there are
certain photon energies at which the effects of real and virtual
particles compensate each other, since the QED vacuum and the plasma
enforce linear polarization of the modes in a mutually orthogonal
direction \citep{lai2003}. The two most relevant of these are at the
vacuum resonance, $E_\mathrm{V}$, which depends on $n_\el$ and
$\vec{B}$ \citep[see, e.g.,][]{ho2003}, and the cyclotron resonance,
$E_\cyc$. The transition through these points results in a flip of the
position angle of the polarization ellipse by $\pm90^\circ$ or in the
reversal of the direction of the rotation of the polarization vector
due to the switch in the dominance of the influence of vacuum and
plasma \citep{pavlov1979}. These effects make the classical definition
of the modes introduced in Sect.~\ref{sec:modclas} internally
inconsistent. As discussed by \citet{kirk1980b}, a general definition
of the modes can be made based on the continuity of the refractive
indices of the waves. Figure~\ref{fig:crs} shows the cross section for
magnetic Compton scattering as a function of photon energy,
illustrating the effect of the continuity of the refractive index
across the spectrum. In the case of a discontinuous refractive index,
corresponding to the change of the wave's handedness, the cross
sections of the modes acquire a sharp, resonance-like feature at
$E_\mathrm{V}$ \citep{ventura1979b}. The alternative choice of
continuous refractive indices results in an abrupt change of opacity
\citep[see, e.g.,][]{meszaros1985a}.

The problem of the mode definition becomes clear when one
considers the propagation in an inhomogeneous medium, where a photon
passes, e.g., from a plasma-dominated region where its energy
$E>E_\mathrm{V}$, to a vacuum-dominated one, where $E<E_\mathrm{V}$.
After passing the ``resonant density'' where $E=E_\mathrm{V}$, modes
with $|K|\ll1$ and $|K|\gg1$ will acquire left-hand and right-hand
polarization, respectively. Figure~\ref{fig:modes} (right) illustrates
the behavior of the modes' ellipticity in an inhomogeneous medium. For
a smooth density gradient, at energies above a few~keV it is expected
that the modes evolve adiabatically (i.e., they keep their original
helicity), while they behave non-adiabatically at lower energies
\citep{oezel2003, ho2003}. In general, however, the probability of a
jump of mode characteristics across the resonance has to be considered
\citep{adelsberg2006}. The adiabatic and non-adiabatic cases
correspond to the continuous and discontinuous behavior of the
refractive index, respectively. In the following, we use the notations
``mode~1'' and ``mode~2'' as denoted for these cases in
Fig.~\ref{fig:crs} by $\sigma_{1,2}$.

\section{Modeling and Results}
\label{sec:model}

We use our polarization-dependent radiative transfer code
\texttt{FINRAD}, which takes into account angular and energy
redistribution of photons both in the continuum and in the cyclotron
line to investigate the effect of vacuum polarization. The code is
based on the Feautrier numerical scheme for two polarization modes
\citep{nagel1981b,meszaros1985a}. More details on the implementation
and verification of the code are given by \citet{sokolova-lapa2021}.
We perform each simulation for both choices of the cross section
behavior displayed in Fig.~\ref{fig:crs}. For the inhomogeneous model,
this choice corresponds to the adiabatic and non-adiabatic mode
propagation. For some models we present a comparison with the modes
for pure plasma (hot, but non-relativistic) and pure vacuum. All
models assume a slab-like emission region, with the $\vec{B}$-field
parallel to the surface normal \citep[same as in][]{nagel1981b}. The
slab is characterized by $n_\el$, the constant $\vec{B}$-field value,
and the electron temperature $kT_\el$. The treatment for induced
processes and boundary conditions are the same as in
\citet{sokolova-lapa2021}, i.e., diffusion limit at the bottom of the
slab and no illumination from the top. We describe the spectral shape
with the photon flux
$F_\mathrm{ph}(E)=4\pi\int^{1}_{0}I(E,\theta)\cos(\theta)/E\,\diff\theta$,
where the specific intensity $I(E,\theta)$ is obtained with
\texttt{FINRAD}. The angular dependence of the emission will be
discussed in future studies. Section~\ref{sec:homog} presents the
spectra from several homogeneous models for various $kT_\el$, $n_\el$,
and $B$. In Sect.~\ref{sec:inhomog}, we then construct a simple model
for an accretion mound with an inhomogeneous density profile and show
the resulting spectra and polarization signal.

\subsection{Homogeneous slab}
\label{sec:homog}

\begin{figure}
    \resizebox{\hsize}{!}{\includegraphics{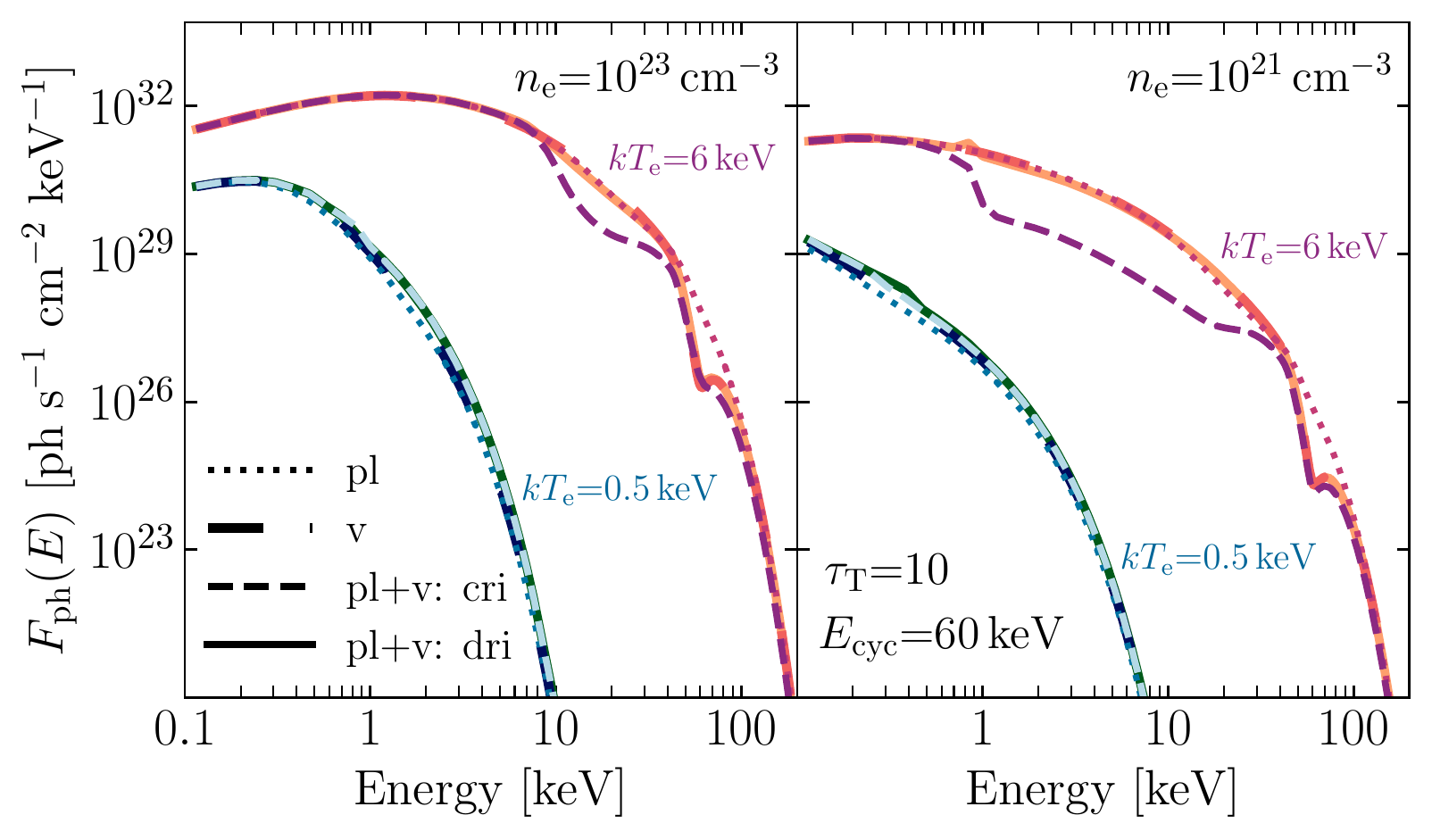}}
    \caption{Flux from a homogeneous slab summed over the two
      polarization modes for the higher (\emph{left}) and lower
      (\emph{right}) electron number density, for
      $B\approx5.2\times10^{12}\gauss$. Different types of
      polarization modes are shown: pure plasma (``pl'', dotted), pure
      vacuum (``v'', thick dashes), modes and modes with
      continuous/discontinuous (``pl${+}$v: cri/dri'', dashed/solid)
      refractive indices. }
    \label{fig:specquad}
\end{figure}

\begin{figure}
    \resizebox{\hsize}{!}{\includegraphics{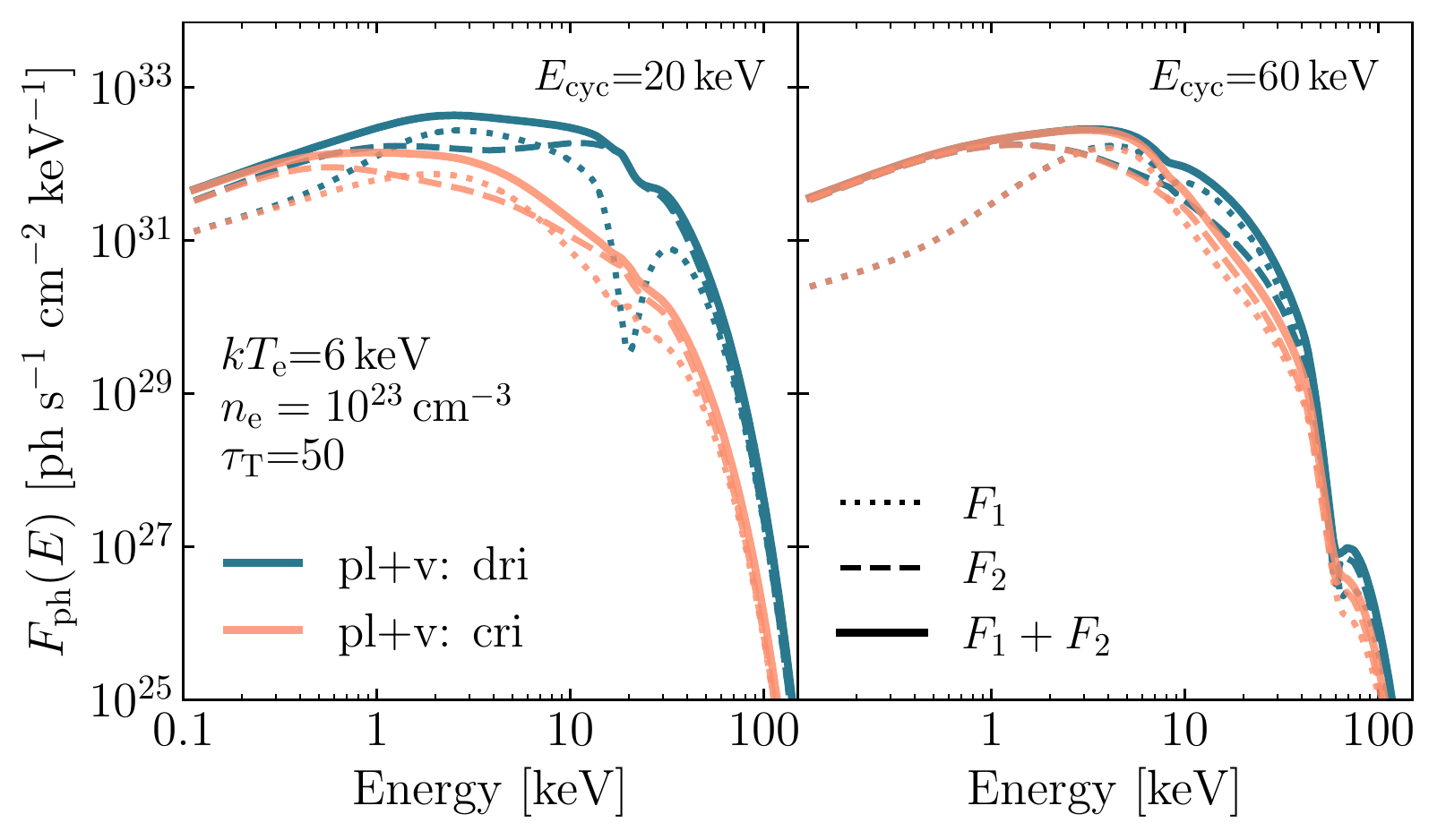}}
    \caption{Emission from a highly magnetized optically thick slab
      for continuous and discontinuous behavior of the refractive indices,
      for two $B$-field values, \emph{left}:
      $B\sim1.7\times10^{12}\gauss$ ($E_\cyc=20\kev$ and
      $E_\mathrm{V}\sim24\kev$) and \emph{right}:
      $B\sim5.2\times10^{12}\gauss$ ($E_\cyc=60\kev$ and
      $E_\mathrm{V}\sim8\kev$). }
    \label{fig:spec_ecyc}
\end{figure}

Figure~\ref{fig:specquad} illustrates the influence of vacuum
polarization on the total spectra of the emission from the homogeneous
slab, and compares the spectra obtained for the pure plasma and
the vacuum cases. The choice of modes for the low-temperature case
($kT_\el=0.5\kev$, spectra with lower flux) has a very minor effect on
the spectral shape. The Comptonized spectra for the hotter plasma,
$kT_\el=6\kev$, illustrate the strong influence of the cyclotron
resonance near the cyclotron energy $E_\cyc=60\kev$ for all cases
except for the pure plasma spectra. The latter are dominated by mode~2
(in this case, the classically-defined ordinary mode), whose weak
resonance is introduced by the Doppler effect on the thermal electrons
and then washed out by angle averaging. The spectra for the case
of the discontinuous refractive indices closely follow the pure
vacuum ones, except for a
small region near the vacuum resonance (the polarization signal,
however, is different, see Sect.~\ref{sec:inhomog}). The spectral
shape for the case of the continuous refractive indices differs
significantly from the others.
There is a noticeable depression of the continuum above the vacuum
resonance energy -- vacuum polarization effectively reduces the total
optical depth of the emitting region. \citet{lai2002} noted similar
behavior in their studies of photon propagation in magnetar-like
fields with $B\sim10^{14}\gauss$. When the vacuum resonance is located
closer to the cyclotron line, it can mimic an additional line-like
feature. For the cases presented in Fig.~\ref{fig:specquad},
$E_\mathrm{V}<E_\mathrm{cyc}$. Figure~\ref{fig:spec_ecyc} (left)
illustrates the case when $E_\mathrm{V}>E_\mathrm{cyc}$ and shows
contributions of the two normal modes. Here, the cyclotron line is
significantly suppressed. The case of the continuous refractive
indices is the most interesting
one, as none of the modes exhibit a strong cyclotron resonance. The
case of a higher $B$-field shown in Fig.~\ref{fig:spec_ecyc} (right)
($E_\mathrm{V}\ll E_\mathrm{cyc}$) yields a more complex continuum
shape.
%

\subsection{Inhomogeneous accretion mound}
\label{sec:inhomog}

In an inhomogeneous medium, the radiation field encounters the vacuum
resonance at a continuum of energies. To understand how this effect
contributes to the total spectrum, we create a simplified model of an
accretion mound confined to the polar cap of a neutron star. The mound
is represented by a slab of magnetized isothermal plasma with
$kT_\el=5\kev$ and $E_\cyc=40\kev$, a height of 500\,m and a radius of
300\,m. We assume that the plasma is accreted at a mass-accretion rate
$\dot{M}=10^{17}\gs$ and that the accretion flow is mainly decelerated
higher up, above the emission region. The velocity on the top of the
mound is set to the $5\%$ of the free-fall velocity near the surface,
$v_\mathrm{ff}$, and decreases linearly within the mound to
$10^{-3} v_\mathrm{ff}$ at the bottom, such that the density
distribution within the mound increases from
${\sim}10^{22}\,\mathrm{cm}^{-3}$ at the top to
${\sim}10^{24}\,\mathrm{cm}^{-3}$ at the bottom.

\begin{figure}
\centering
    \resizebox{\hsize}{!}{\includegraphics{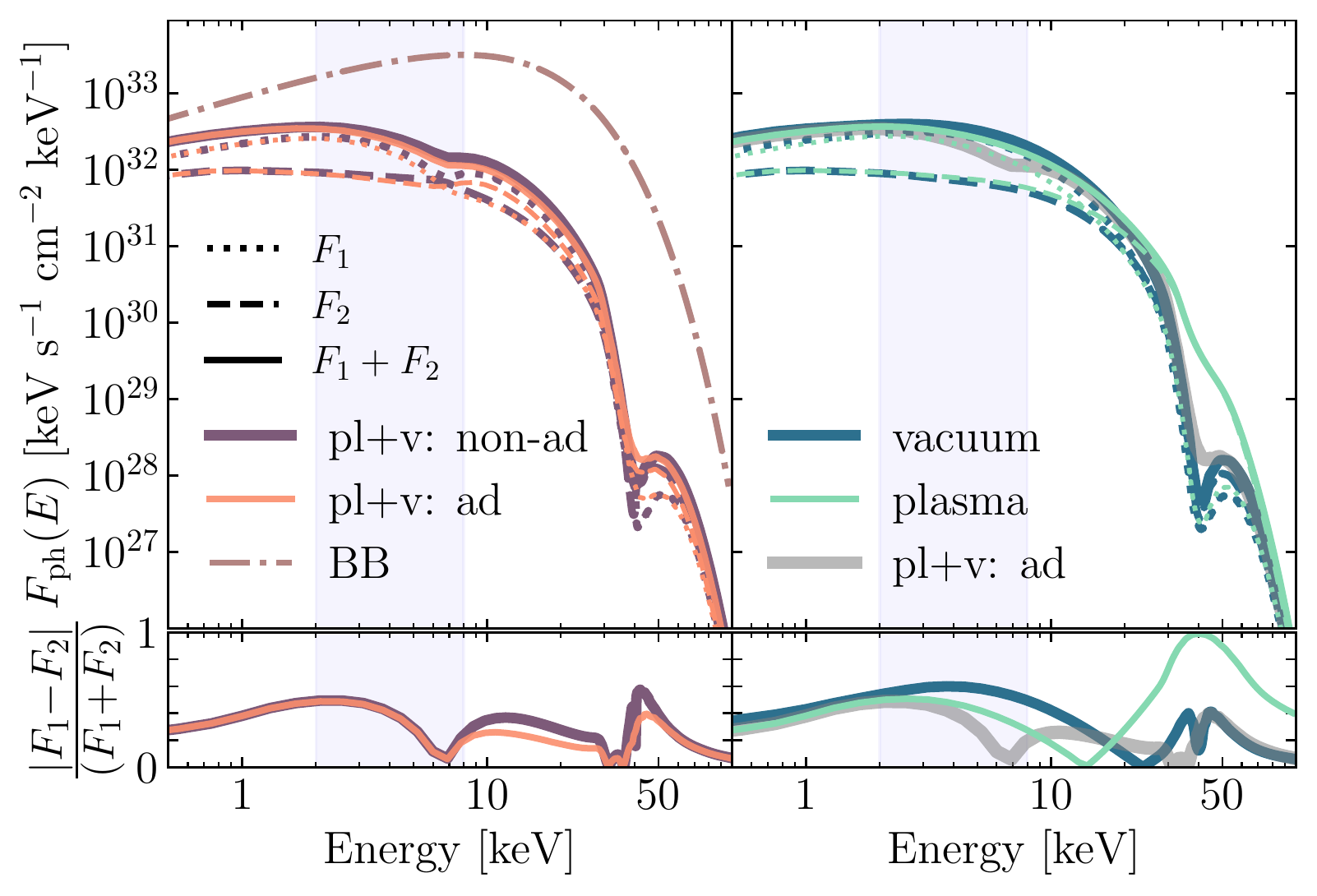}}
    \caption{\emph{Top:} Flux from an inhomogeneous mound with
      $kT_\el=5\kev$ and $B\sim3.6\times10^{12}\gauss$
      ($E_\cyc=40\kev$) for the cases including the plasma and vacuum
      polarization effects for non-adiabatic,``non-ad'' and adiabatic, ``ad'',
      mode propagation (\emph{left}) and the pure plasma and
      vacuum cases (\emph{right}). The left-top panel shows the
      corresponding blackbody spectrum (``BB''). \emph{Bottom:} The
      corresponding degree of polarization. The energy range
      accessible by the Imaging X-ray Polarimetry Explorer (IXPE) is
      shown by the light-blue transparent region.}
    \label{fig:spec_inhom}
\end{figure}
Figure~\ref{fig:spec_inhom} shows the total flux for the previously
discussed types of polarization modes. Here, the adiabatic and
non-adiabatic cases (left panel) are similar, both exhibiting an
imprint of the vacuum resonance at ${\sim}6.5\kev$ from the most
tenuous top layer of the slab. For adiabatic propagation the total
flux is slightly suppressed above $E_\mathrm{V}$, with the cyclotron
line core becoming noticeably more shallow. Due to the absence of the
vacuum resonance, the pure plasma and vacuum cases (right panel)
naturally lack the additional complexity of the spectra.

For both cases, the polarization degree (Fig.~\ref{fig:spec_inhom},
bottom) reaches a maximum of
${\sim}50\%$\footnote{For the angle-resolved spectra of our
  calculations, not presented here, the polarization degree reaches a
  maximum of ${\sim}65\%$ in the continuum, at
  $\theta\sim70^{\circ}$}. It is overall lower than predicted by
the accretion column model of \citet{caiazzo2021a} for the
emission in the neutron star rest frame.
This difference is probably due to radiative propagation effects that
were omitted by these authors, such as redistribution during
scattering, which are enhanced by the presence of the hot plasma. The
degree of polarization in the pure vacuum case is higher at
intermediate energies; for the plasma case it reaches almost 100\%
near the cyclotron line.

\section{Discussion and Conclusions}
\label{sec:concl}

In this work we presented the results of radiative transfer
calculations in the hot Comptonizing medium in the vicinity of
accreting neutron stars with $B\sim10^{12}\gauss$, including the
effect of vacuum polarization. We found that the choice of
polarization modes and the vacuum resonance can potentially strongly
influence the emitted spectrum. Together with angle- and
energy-dependent magnetic Comptonization, those polarization effects
result in a complex continuum shape and can alter the cyclotron line.
The natural switch in the dominance of the two polarization modes in
the total flux across the spectrum produces excesses and dips on top
of the power-law-like continuum. While this phenomenon is not specific
for vacuum polarization, it is enhanced by it due to the additional
interplay between the modes. The vacuum resonance itself can also
introduce a weak dip in the spectra, mimicking the cyclotron line, as
shown for the inhomogeneous model. For the conditions studied here, we
find that pure vacuum normal modes provide a closer spectral shape to
the mixed (magnetized plasma and polarized vacuum) case than pure
plasma modes.

The combination of these effects may also be able to
explain several complexities observed in the spectra of accreting
neutron stars in HMXBs, such as the ``10-keV'' feature, or the
two-component-like continua. Stand-alone absorption line-like features
at intermediate energies \citep[see, e.g.,][for 4U~1901+03 and
KS~1947+300]{reig2016, doroshenko2020a} are potential candidates
for the mode interchange regions. Vacuum polarization can also result
in a significant suppression of the cyclotron line when
$E_\mathrm{V}>E_\cyc$, potentially explaining the missing cyclotron
lines in the spectra of some sources.

Concerning the polarization degree, our results for a simplified
model of the accretion mound show a maximum at ${\sim}50\%$ in
the continuum. The vacuum resonance results in a broad region of
depolarization. A detailed understanding of the angular redistribution
is important to draw any further conclusions, which will be addressed
in the future work, along with the study of emission beaming and pulse
profile formation under the influence of light bending. We expect that
angle-dependent beaming and the propagation of the X-rays in the
magnetosphere, will only lower the observable degree of polarization,
bringing the total value closer to the low values observed recently by
IXPE \citep[see, e.g.,][]{tsygankov2022,doroshenko2022}. The presented model is a first step towards a more realistic description
of the radiation field in the accretion channel, which ideally
should include partial mode conversion, higher relativistic
corrections, and consistent simulation of the structure of the
accretion mound, as well as the influence of the falling flow.

\begin{acknowledgements}
  This research has been funded by DFG grants WI1860/11-2 and
  WI1860/19-1. The work originates in part from the text of the thesis
  draft by ESL. We are grateful to Aafia Zainab, Philipp Thalhammer,
  and Katrin Berger for their valuable comments. The
  authors also thank the XMAG collaboration and all participants of
  the workshop ``X-ray Tracking of Magnetic Field Geometries in
  Accreting X-ray Pulsars'', who supported fascinating discussions and
  raised new inspiring questions for this research.
\end{acknowledgements}


\end{document}